\documentclass[aip,pop,reprint,floatfix]{revtex4-1}

\usepackage{graphicx}
\usepackage{amssymb}
\usepackage{amsmath}
\usepackage{xspace}
\usepackage{setspace}
\usepackage{tabularx}
\usepackage[colorlinks=true]{hyperref}
\usepackage[sort&compress]{natbib}
\setcitestyle{sort&compress,maxcitenames=3}

\newcommand{\rfr}[1]{\cite{#1}}
\ifdefined\qed
	\renewcommand{\qed}{\ensuremath{\hspace{1em}_{\blacksquare}}\xspace}
\else
	\newcommand{\qed}{\ensuremath{\hspace{1em}_{\blacksquare}}\xspace}
\fi
\newcommand{\eq}[2][]{\begin{equation}\label{eqn:#1}\begin{split}#2\end{split}\end{equation}}
\newcommand{\eqr}[1]{Eq.~(\ref{eqn:#1})}
\newcommand{\fg}[4][]{\begin{figure}[h!tbp]\begin{center}\includegraphics[width=#4mm]{#3}\caption{#1\label{fig:#2}}\end{center}\end{figure}}
\newcommand{\fgr}[1]{Fig.~\ref{fig:#1}}

\newcommand{\del}{\partial}
\newcommand{\deldel}[2][]{\ensuremath{\frac{\del #1}{\del #2}}\xspace}

\newcommand{\deldt}[1][]{\deldel[#1]{t}}

\newcommand{\he}[1][]{\ensuremath{^{\textrm{#1}}\textrm{He}}\xspace}
\newcommand{\prot}[1][]{\ensuremath{\textrm{p}}\xspace}
\newcommand{\deut}[1][]{\ensuremath{\textrm{D}}\xspace}
\newcommand{\trit}[1][]{\ensuremath{\textrm{T}}\xspace}

\renewcommand{\v}{\ensuremath{\boldsymbol{v}}\xspace}

\newcommand{\erf}{\textrm{erf}\xspace}

\makeatletter
\newcommand\etc{\textit{etc}\@ifnextchar.{}{\textit{.}\xspace}}
\newcommand\versus{\textit{versus}\xspace}
\newcommand\eg{\textit{e.g}\@ifnextchar.{}{\textit{.}\xspace}}
\newcommand\etal{\textit{et~al}\@ifnextchar.{}{\textit{.}\xspace}}
\newcommand\via{\textit{via}\xspace}
\makeatother

\newcommand{\figwidth}{80}

\begin{document}


\title{Particle-in-cell studies of fast-ion slowing-down rates in cool tenuous magnetized plasma} 



\author{Eugene S. Evans}
\email[]{eevans@pppl.gov}
\affiliation{Princeton Plasma Physics Laboratory}

\author{Samuel A. Cohen}
\affiliation{Princeton Plasma Physics Laboratory}

\author{Dale R. Welch}
\affiliation{Voss Scientific}


\date{\today}

\begin{abstract}
We report on 3D-3V particle-in-cell simulations of fast-ion energy-loss rates in cold, weakly-magnetized, weakly-coupled plasma where the electron gyroradius, $\rho_{e}$, is comparable to or less than the Debye length, $\lambda_{De}$, and the fast-ion velocity exceeds the electron thermal velocity, a regime in which the electron response may be impeded. These simulations use explicit algorithms, spatially resolve $\rho_{e}$ and $\lambda_{De}$, and temporally resolve  the electron cyclotron and plasma frequencies. For mono-energetic dilute fast ions with isotropic velocity distributions, these scaling studies of the slowing-down time, $\tau_{s}$, \versus fast-ion charge are in agreement with unmagnetized slowing-down theory; with an applied magnetic field, no consistent anisotropy between $\tau_{s}$ in the cross-field and field-parallel directions could be resolved. Scaling the fast-ion charge is confirmed as a viable way to reduce the required computational time for each simulation. The implications of these slowing down processes are described for one magnetic-confinement fusion concept, the small, advanced-fuel, field-reversed configuration device.
\end{abstract}

\pacs{}

\maketitle 

\section{Introduction}

Ions with velocities greater than those of most particles in the surrounding plasma are found across all scales of plasma physics, from high energy cosmic rays passing through the interstellar medium\rfr{everett_interaction_2011} to heavy ions propagating into solid targets\rfr{geissel_experimental_2002}. Understanding the fast-ion energy-transfer processes in fusion plasmas with an energetic component has also long been a topic of considerable interest. How rapidly fast ions slow down is important to the dynamics of and energy balance in their background plasmas. Research for magnetic fusion energy (MFE) has examined situations as diverse as intense circulating beams formed by ionization of energetic neutral beams used for plasma heating\rfr{stix_heating_1972} to tenuous isotropic mono-energetic distributions of fusion-produced alphas\rfr{hurricane_inertially_2016}. The hot core plasma of fusion devices has been particularly well-studied. There, the magnetic field has little effect on slowing down: fast-ion velocities are much less than the thermal velocity of the background electrons and the electron gyroradius, $\rho_{e}$, is large compared to the Debye length, $\lambda_{De}$.

This work looks at a less-well-studied (still weakly-coupled) regime where the fast-ion velocity, $v_{fi}$, is comparable to the electron thermal velocity, $v_{th,e}$, and  $\eta\equiv\lambda_{De}/\rho_{e} \lesssim 2$ (see \fgr{parameter-phase-space}). Both of these conditions might be expected to inhibit the electron response to the energetic ion, hence reduce the efficiency of electron drag on the ions and the slowing-down rate.  Such conditions are present in the edge plasma of many MFE devices, though are considered relatively unimportant in large devices whose plasma radii, $a$, are much greater than $\rho_{i}  $. We will describe why they can be important to power and ash exhaust in small MFE devices where $a \sim \rho_{i} $. Note that we restrict attention to cases where the energetic particle density is much less than the background plasma density and do not include transit-time-type resonances, {\it e.g.,} excluding streaming\rfr{davidson_survey_2009} and GAE-type instabilities\rfr{appert_excitation_1982}, respectively (either could increase the energy loss rate).

\fg[The blue region represents the combination of parameters under consideration in this paper. Red corresponds to classical (unmagnetized) slowing down, green is highly-coupled, highly-magnetized slowing down\rfr{dubin_parallel_2014}, orange is ions slowing down in matter\rfr{geissel_experimental_2002}, purple is high-energy cosmic rays slowing down in diffuse plasma\rfr{everett_interaction_2011}. $\Gamma$, $\Omega_{ce}$, $\omega_{pe}$, $v_{fi}$, and $v_{th,e}$ are the plasma coupling parameter, the electron cyclotron frequency, the electron plasma frequency, the fast ion velocity, and the electron thermal velocity, respectively.]{parameter-phase-space}{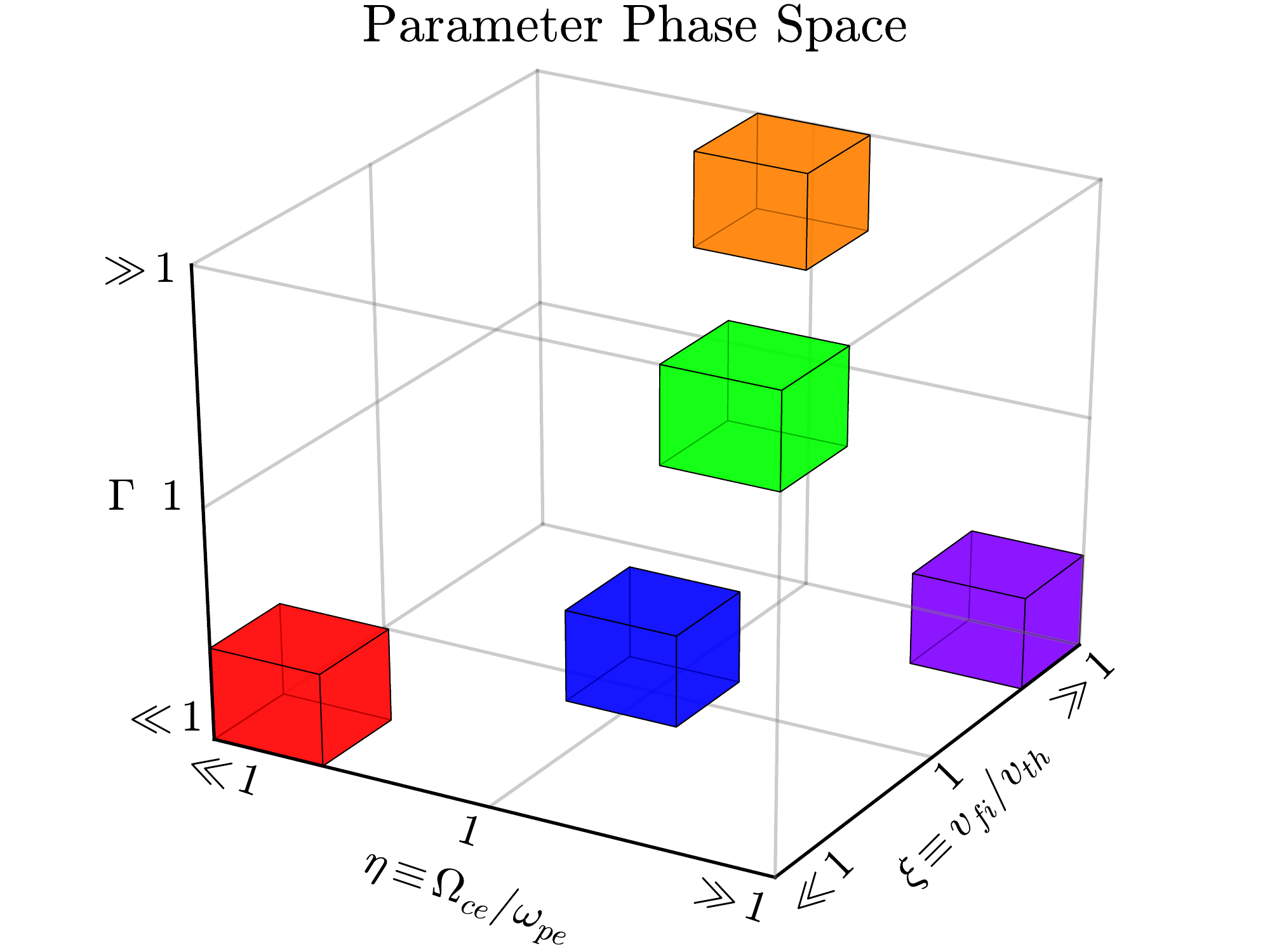}{80}

Rostoker \rfr{rostoker_n._kinetic_1960}, among others\rfr{rostoker_test_1960}, derived a Fokker-Planck operator incorporating the effects of an external magnetic field; only the asymptotic cases of $\eta\ll 1$ and $\eta\gg 1$ were considered in detail. Much of the work building on these early results has tended toward analysis of strongly-magnetized systems ($\eta\gg 1$), with either very large magnetic fields \rfr{ware_electron_1989,dubin_parallel_2014,affolter_first_2016} or very low temperature plasmas \rfr{oneil_collision_1983}. The results obtained are not applicable to the case of one unmagnetized projectile species (with trajectories unperturbed by the magnetic field) scattering on a weakly magnetized field species, since the field particles are only approximately confined to move along field lines. Analytical work has been done on slowing down rates in weakly magnetized systems, but explicit calculations of slowing down times have been limited to the subthermal ($v\ll v_{th,e}$) regime, e.g. Hassan and Watson\rfr{hassan_magnetized_1977}. Hassan and Watson found a correction to the field-parallel slowing down rate of -5.2\%, and an asymmetry in the cross-field \versus field-parallel slowing down rates of -2\% (slower cross-field slowing down) for $\eta=2$ and $v=1.34 v_{th,e}$ (see \fgr{hw-diff}). One way forward in examining how best to extend these various analytical theories to our regime of interest where $\eta\sim 1$ was to study the issue \via direct simulation. Recent work in direct simulations of slowing down in a molecular dynamics system \rfr{grabowski_molecular_2013} and others \rfr{robiche_fast_2010,rose_particle--cell_2009} suggest that the problem may now be computationally feasible using PIC codes.

\section{Slowing Down Theory}

Classical slowing down theory was first formulated by Rosenbluth, \etal, in 1957\rfr{rosenbluth_fokker-planck_1957}, followed by Rostoker's extension of the theory to a homogenous background magnetic field in 1960\rfr{rostoker_n._kinetic_1960}, using the BBGKY theory approach. Here we briefly discuss nonmagnetized classical slowing down theory in order to emphasize a few of its scaling properties in the context of PIC codes and formulate an appropriate model for comparison with subsequent simulations.

To avoid confusion when comparing simulation results with theory, we use the following definition of the (energy) slowing-down time:

\eq[slowing-down-def]{\frac{1}{\tau_{s}}=-\frac{\deldt[W]}{W}}

We start from the the Fokker-Planck equation, following \textsc{Bellan}'s treatment\rfr{bellan_fundamentals_2006} of \textsc{Rosenbluth}. We use the first velocity moment, and assume that the fast ions, (the test particles,  subscript $T$), are mono-energetic and the background (subscript $F$, for field particles, both electrons and ions) particles are  Maxwellian, obtaining (MKS units)

\eq[classical]{\deldt[u]=\sum\limits_{F}\frac{n_{F}q_{T}^{2}q_{F}^{2}\ln\Lambda}{4\pi\epsilon_{0}^{2}m_{T}^{2}}\frac{m_{T}}{\mu_{F}}\frac{m_{F}}{2\kappa T_{F}}\left(\deldel{\xi_{F}}\left(\xi_{F}^{-1}\erf\left(\xi_{F}\right)\right)\right)}

\noindent where $\xi_{F}=\sqrt{\frac{m_{F}}{2\kappa T_{F}}}u$ is the ratio between the fast ion velocity $u$ and the field particle thermal velocity, while $m_{f}$ and $T_{f}$ are the mass and temperature of the field particles, $\erf$ is the error function, and $\mu_{F}$ is the reduced mass. One of the approximations made by a PIC code is that each macroparticle in the simulation represents some number of real particles, denoted by the clumping factor $\zeta$. The clumping factor modifies the properties of a macroparticle in the following way:

\eq[macroparticle-scaling]{n&\rightarrow n/\zeta\\m&\rightarrow\zeta m\\q&\rightarrow\zeta q\\T&\rightarrow\zeta T\\\ln\Lambda&\rightarrow\ln\left(\Lambda/\zeta\right)}

Note that all velocities and relevant frequencies are preserved by \eqr{macroparticle-scaling}, as well as the Debye length. Incorporating these transformations into \eqr{classical}, we have

\eq[macro-scaled]{\deldt[u]=\sum\limits_{F}\zeta_{F}\frac{n_{F}q_{T}^{2}q_{F}^{2}\ln\left(\Lambda/\zeta_{F}\right)}{4\pi\epsilon_{0}^{2}m_{T}^{2}}\left(1+\frac{\zeta_{T}m_{T}}{\zeta_{F}m_{F}}\right)\frac{m_{F}}{2\kappa T_{F}}\times\\\left(\deldel{\xi_{F}}\left(\xi_{F}^{-1}\erf\left(\xi_{F}\right)\right)\right)}

For the plasma conditions we are interested in (namely, $\v_{T} >> \v_{th,i}$), the contribution of the background ions to the slowing down force will be negligible, so we restrict \eqr{macro-scaled} to electrons only (field particles will henceforth be denoted with subscript $e$). We now consider the quantity $\left(1+\frac{\zeta_{T}m_{T}}{\zeta_{e}m_{e}}\right)$ and note that if $\zeta_{T}\gtrsim\zeta_{e}$ and $m_{T}\sim m_{p}$, then $\left(1+\frac{\zeta_{T}m_{T}}{\zeta_{e}m_{e}}\right)\approx \frac{\zeta_{T}m_{T}}{\zeta_{e}m_{e}}$, simplifying \eqr{macro-scaled} further:

\eq[macro-scaled-electrons]{\deldt[u]=\frac{n_{e}e^{2}\ln\left(\Lambda/\zeta_{e}\right)}{4\pi\epsilon_{0}^{2}(2\kappa T_{e})}\frac{\zeta_{T}q_{T}^2}{m_{T}}\left(\deldel{\xi_{e}}\left(\xi_{e}^{-1}\erf\left(\xi_{e}\right)\right)\right)}

Note that \eqr{macro-scaled-electrons} is independent of the electron $\zeta$. Using the asymptotic approximations of $\deldel{\xi_{e}}\left(\xi_{e}^{-1}\erf\left(\xi_{e}\right)\right)$ for large and small $\xi_{e}$, we have:

\eq[electrons-asymp]{\deldt[u]=-\frac{n_{e}e^{2}\ln\left(\Lambda/\zeta_{e}\right)}{4\pi\epsilon_{0}^{2}(2\kappa T_{e})}\frac{\zeta_{T}q_{T}^2}{m_{T}}\begin{cases}\frac{4\xi_{e}}{3\sqrt{\pi}},&\ \xi_{e}\ll 1\\\frac{1}{\xi_{e}^2},&\ \xi_{e}\gg 1\end{cases}\\
=-\frac{n_{e}e^{2}\ln\left(\Lambda/\zeta_{e}\right)}{4\pi\epsilon_{0}^{2}}\frac{\zeta_{T}q_{T}^2}{m_{T}}\begin{cases}\frac{4m_{e}^{1/2}u}{3\sqrt{\pi}(2\kappa T_{e})^{3/2}},&\ \xi_{e}\ll 1\\\frac{1}{m_{e}u^{2}},&\ \xi_{e}\gg 1\end{cases}}

Converting to energy, where $W=\frac{1}{2}\zeta_{T}m_{T}u^{2}$, and substituting into \eqr{electrons-asymp}, we obtain the following two expressions:

\eq[tau-energy-general]{\frac{1}{\tau_{s}}=\frac{n_{e}e^{2}\ln\left(\Lambda/\zeta_{e}\right)}{4\pi\epsilon_{0}^{2}}\frac{\zeta_{T}q_{T}^2}{m_{T}}\begin{cases}\frac{8m_{e}^{1/2}}{3\sqrt{\pi}(2\kappa T_{e})^{3/2}},&\ \xi_{e}\ll 1\\\frac{\zeta_{T}^{3/2}m_{T}^{3/2}}{\sqrt{2}m_{e}W^{3/2}},&\ \xi_{e}\gg 1\end{cases}}

For the $\xi_{e}\ll 1$ case, we have:

\eq[tau-energy-subthermal]{\frac{1}{\tau_{s}}&=\frac{\zeta_{T}q_{T}^{2}}{m_{T}}\left(\frac{n_{e}}{(\kappa T_{e})^{3/2}}\right)\left(\frac{m_{e}^{1/2}e^{2}\ln\left(\Lambda/\zeta_{e}\right)}{3\sqrt{2}\pi^{3/2}\epsilon_{0}^{2}}\right),\\&W(t)=W_{0}e^{-t/\tau_{s}}}

As our particular simulations involve projectile particles with the mass of \he[4], we finally specialize \eqr{tau-energy-subthermal} to fast ions with $m_{T}=4m_{p}$ and $q_{T}=Z_{fi}e$:

\eq[alpha-electron-energy]{\frac{1}{\tau_{s}}&=\zeta_{fi}Z_{fi}^{2}\left(\frac{n_{e}}{(\kappa T_{e})^{3/2}}\right)\left(\frac{m_{e}^{1/2}e^{4}\ln\left(\Lambda/\zeta_{e}\right)}{12\sqrt{2}\pi^{3/2}\epsilon_{0}^{2}m_{p}}\right),\\&W(t)=W_{0}e^{-t/\tau_{s}}}

This simple model will be used as the basis for comparison with simulation results (both the subthermal benchmarks and the non-subthermal scaling studies). In particular, it is the $Z_{fi}^{2}$ factor that will allow for orders-of-magnitude acceleration in slowing-down-time simulations when $Z_{fi}$ is scaled up.

\textsc{Baldwin and Watson} derived ``magnetized Rosenbluth potentials" for the case of a magnetized plasma where the ion Larmor radius is much larger than the Debye length (so straight-line ion trajectories are assumed), but the electron Larmor radius can be comparable to or less than the Debye length. When evaluated for the case of fast ions interacting with a Maxwellian background when $\xi_{e}\lesssim 1$\rfr{hassan_magnetized_1977}, this modified $\tilde{h}$ potential has the following form:

\eq[hw-h-maxwellian]{\tilde{h}=\frac{m_{i}}{m_{e}}\Gamma^{ie}\left[\frac{\Phi(v/v_{th,e})}{v}\ln\left(\frac{\Lambda_{0}}{\sqrt{1+\eta^{2}}}\right)\right.\\+\left.\frac{e^{-r^2+s^2}}{\pi^{1/2}v_{th,e}}s^{2}\ln(1+\eta^{2}/2)^{1/2}\right.\times\\\left.\left(K_{1}-K_{0}+2r^{2}\left[K_{0}+2s^{2}(K_{0}-K_{1})\right]\right)\right]}

\noindent($K_{0}$ and $K_{1}$ are modified Bessel functions of the second kind, with arguments of $s^{2}$, $\Phi$ is the error function, $r$ is $v_{||}/v_{th,e}$, $s$ is $v_{\perp}/v_{th,e}\sqrt{2}$, and $\Gamma^{ie}$ is $4\pi\left(Z_{fi}e^2/4\pi\epsilon_{0}m_{fi}\right)^{2}$)

Mathematica was used to evaluate the friction coefficients using \eqr{hw-h-maxwellian}; the difference between the cross-field ($v_{||}=0$) and field-parallel ($v_{\perp}=0$) slowing down rates is plotted as a function of $\zeta_{T}$ and $\eta$ in \fgr{hw-diff}. For $\eta=2$ and $\xi_{T}=1.34$ (the case of a 1~MeV triton), the cross-field slowing down rate ($\sim 1/\tau_{\perp}$) is 2-3\% smaller than the field-parallel slowing down rate ($\sim 1/\tau_{||}$), a small correction to the unmagnetized case. This is consistent with the intuition that, as the system is only weakly magnetized, electron mobility near fast ions is not severely impacted.

\fg[Percent difference between $\tau_{\perp}$ and $\tau_{||}$ (normalized to $\tau_{||}$); positive values indicate longer cross-field slowing times ($\tau_{\perp}>\tau_{||}$). ]{hw-diff}{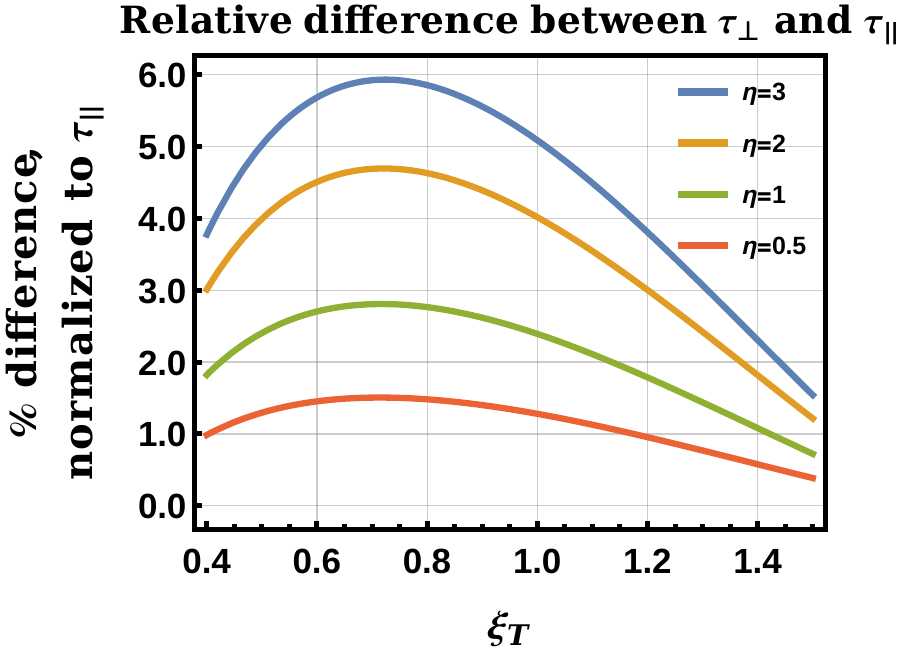}{\figwidth}

To see this, consider a 1~MeV triton moving across a 5~T background magnetic field and a counter-propagating $100$~eV electron, with some perpendicular separation $d<\lambda_{De}$. If $d=\lambda_{De}/\sqrt{2}$, then the electron feels the effect of the triton for $\sim 0.075$~ns, and over the course of the interaction moves $\sim 0.03$~nm toward the triton (a very small fraction of its gyroradius) and gains $\sim 3\times 10^{-29}$~J. For a background plasma density of $10^{14}$/cc, a stream of such counter-propagating electrons could remove 1~MeV from the fast ion in $\sim 3$~ms.

\section{Simulations}

\subsection{Setup}

The PIC code used for these simulations is LSP\rfr{welch_simulation_2001}, as has been used in previous efforts in this project, including initial magnetized slowing down simulations in 2D\rfr{creely_particle--cell_2012}, detailed studies of the effects of macroparticle clumping and collision models in LSP on slowing down \rfr{kolmes_applications_2014}, and two-stream instability studies in 2D showing enhanced slowing down for beams\rfr{paul_2-d_2013}. All runs were performed using an energy-conserving explicit particle pusher\rfr{welch_simulation_2001} and an explicit field solver; in the absence of fast ions, LSP conserves energy to better than 1 part in $10^{6}$ over the course of a 30 ns simulation. As a consequence of the use of explicit algorithms, an appropriate cell size was chosen to resolve $\lambda_{De}$ (as well as $\rho_{e}$ when including an external magnetic field), typically by a factor of 5 to 7. Note that although LSP includes a number of different subgrid collision models, we disabled them to avoid double counting the collisions, a point we shall return to later in this paper. A collisional model is not appropriate for these simulations because ionization and fluid effects are unimportant at the temperatures and densities of these weakly coupled plasmas and the number of particles in the well-resolved Debye sphere is large. Thus, the frictional drag that the fast ions experience is entirely due to collective effects and communicated \via the electric field grid.

One consequence of descretizing space is that the simulation Coulomb logarithm takes on a different value than it would as calculated from the plasma parameters in use; specifically, the lower bound on the integral is the cell size rather than the interparticle spacing. This simulation Coulomb logarithm was computed as

\eq[loglambdasim]{\ln\Lambda_{s}=\ln\left(\lambda_{De}/(\textrm{cell size}/2)\right)}

Here, the factor of 2 results from the use of higher-order particle shapes in LSP (to reduce particle self-force). The effects of this change were investigated by varying the cell size.

All frequencies (notably $\omega_{pe}$ and $\Omega_{e}$) were resolved to at least 1 part in 100, and at least 60 periods of $\omega_{pe}$ were included in each simulation; the combination of these conditions ensured that the background electron distribution stayed Maxwellian. When a background magnetic field was used, it was uniform, and oriented in the +Z direction.

A characteristic simulation consisted of a box in 3D cartesian space with a side length of about  $\sim5 \lambda_{De}$ and periodic boundary conditions. Each cell size was no larger than one-seventh the Debye length, and with 64 particles per cell of each background species, there were at least $10^5$ electron macroparticles in each Debye sphere. The electron density in these simulations was chosen to be $10^{12}$/cc (unclumped), in order to relax some of the computational constraints, while the applied magnetic field was chosen to give a range of $\eta$ between 0 and $\lesssim 2$. This placed the simulation in the same dimensionless parameter space (of $\eta \lesssim 2$) as 100x higher density and 10x higher magnetic field.  Finally, both background species were created uniform in space but Maxwellian in energy, with the electrons at 100 eV and the ions at 1 eV.

In order to maintain charge neutrality and avoid spurious charge buildup, the fast ions were injected with a contingent of neutralizing electrons with the same temperature as those in the background plasma. Most commonly, the fast ions were injected with isotropically directed velocites and homogenous in space, with monoenergetic starting energies. However, non-isotropic distributions were also used to explicitly study field-parallel \versus cross-field slowing down rates. In order to not perturb the background plasma, the density of fast ions was kept to a very small fraction of the background, approximately $10^{6}$/cc, or about 1 fast ion per Debye sphere, a percentage that is comparable to MFE-device scrape-off layers. Also, in order to avoid nonlinear effects due to high-Z projectiles\rfr{zwicknagel_nonlinear_1998}, the fast ion charge was limited to $Z=2000e$, giving any fast ion at most 1\% of the charge in a given Debye sphere.

\subsection{Cluster Performance}

The majority of the simulations were run on NERSC clusters (Hopper, Edison, and Cori), with some performed on PPPL's cluster. Typical job sizes were 192 or 768 processors, with run times from 6 to 24 hours; the restart functionality provided by LSP was used to extend simulations where needed. Computational efficiency was found to be around 2~ps of simulation time per hour of charged time, with slightly increased efficiency for the 192-processor jobs. Overall, a typical simulation consumed 20000-50000 CPU hours. However, computational efficiency declined significantly as additional regions were used. Of the $\sim 2.4$~million CPU hours represented in \fgr{cell-scan-lambda}, $\sim 0.8$~million were consumed just by the 384- and 576-CPU runs.

\section{Results}

\subsection{Subthermal Benchmarking}

We  benchmarked the PIC simulations for fast ions with velocities well below the electron thermal velocity and zero magnetic field, to explore the effects of varying the macroparticle clumping factor and charge. The subthermal regime was chosen to avoid  wake effects from superthermal particles, which would require a much larger simulation volume. The fast ions were injected isotropically and homogeneously with the mass of an alpha particle and either $50$~keV or $100$~keV of energy (before clumping) each.

\fgr{50kev-history} shows a typical time history for a high-Z simulation, along with an exponential fit; while the time histories of individual particles may diverge substantially from each other, the average exhibits good agreement with an exponential fit. Although the simulation is just under 8~ns long, the fast ions experience a drop in energy of approximately half an e-folding (factor of $\sim ~0.85$).

\fg[Representative time history for high-Z fast ions ($n_{e}=10^{12}$/cc, $E_{fi}=50$~keV, $Z_{fi}=2000$, $\zeta_{fi}=10$, $B_{z}=0$~T)]{50kev-history}{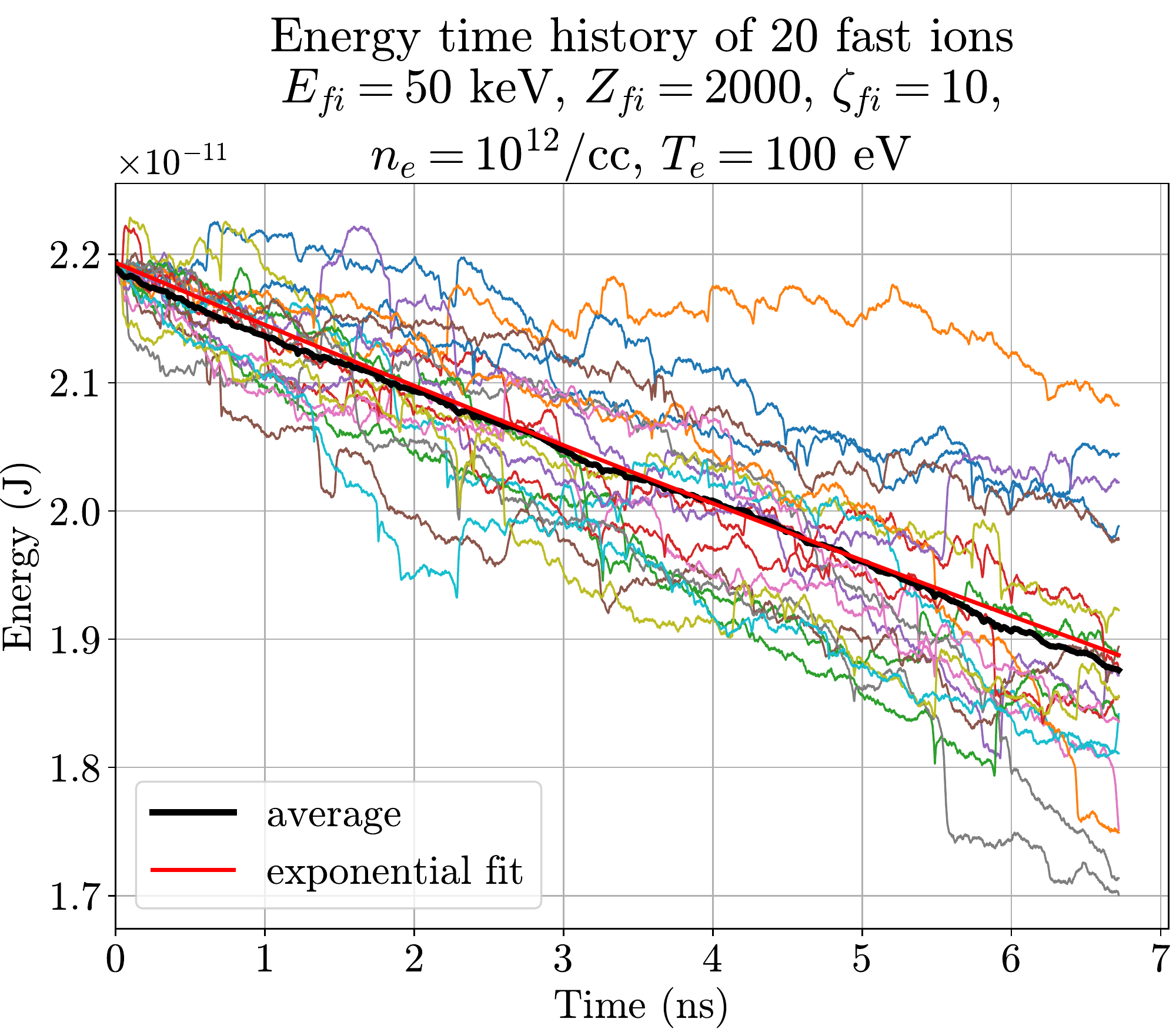}{\figwidth}

In order to match the subthermal model with the simulation data and verify that we were using the correct calculation for the simulation Coulomb logarithm, a scan in cell size was carried out, resulting in \fgr{cell-scan-cells} and \fgr{cell-scan-lambda}. In these runs, the simulation volume was held constant while the number of cells per linear dimension were scanned.

\fg[Time histories for fixed-volume runs with different numbers of cells (per dimension); the 384- and 576-CPU simulations only ran for $\sim 1/\omega_{pe}$.]{cell-scan-cells}{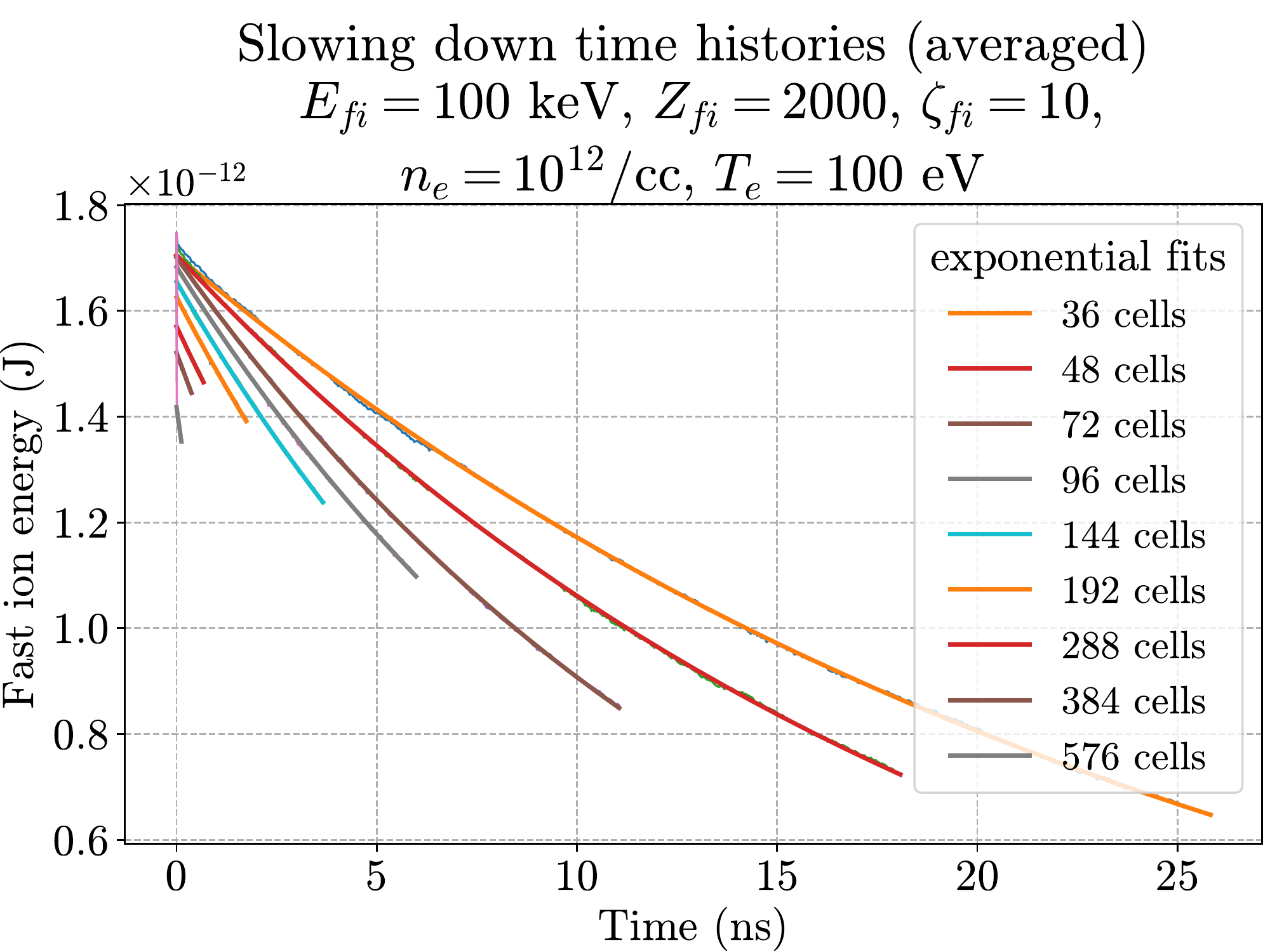}{\figwidth}

\fg[$\tau_{s}$ as a function of simulation $\ln\Lambda$; the subthermal prediction is in red, with the predicted $\tau_{s}$ at $\ln\Lambda=14.7$ indicated. Error bars for the data point at $\ln\Lambda=4$ were established by comparing the behavior of each curve in \fgr{cell-scan-cells} before and after $0.2$~ns.]{cell-scan-lambda}{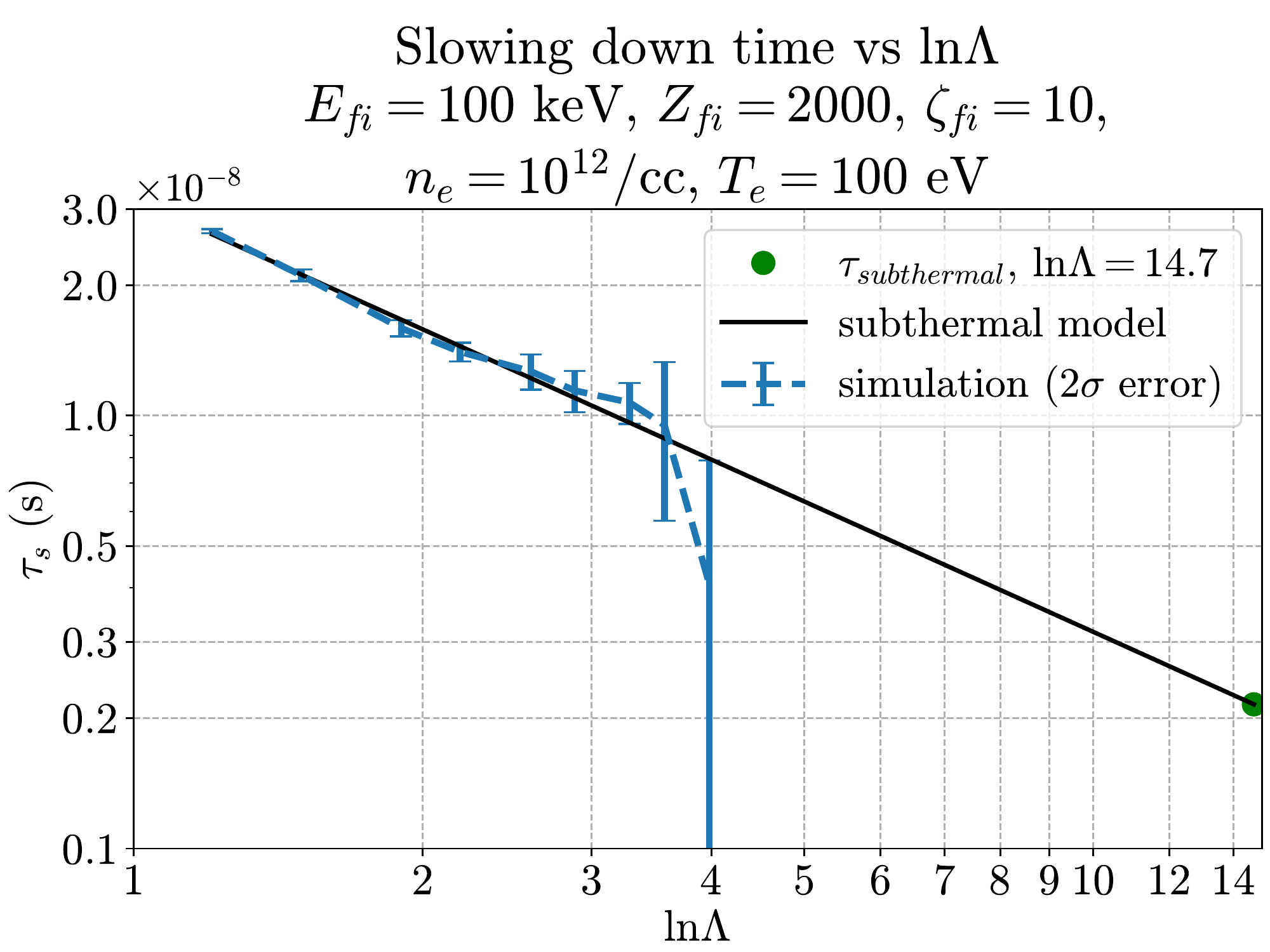}{\figwidth}

Finally, the overall trend in $\tau_{s}$ with $\zeta_{fi}$ and $Z_{fi}$ is shown in \fgr{zzetabscale}; the subthermal prediction includes the simulation $\ln\Lambda$ factor, as determined from \fgr{cell-scan-lambda}.

\fg[$\tau_{s}$ as a function of fast ion macroparticle clumping factor $\zeta_{fi}$ and fast ion charge $Z_{fi}$, with and without a weak background magnetic field. The classical subthermal prediction is in black.]{zzetabscale}{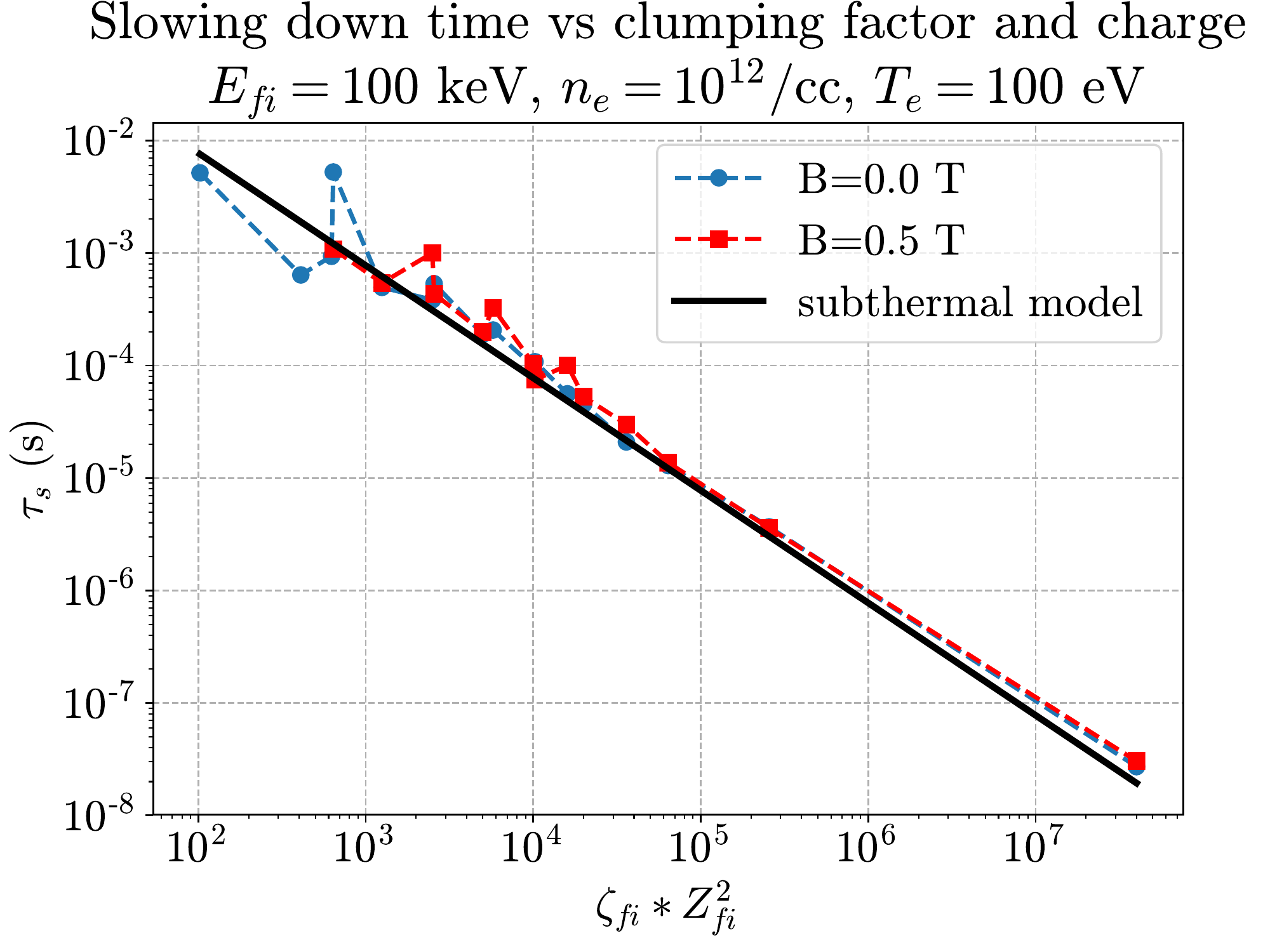}{\figwidth}

\subsection{Scaling with Energy and Magnetic Field}

The simulations in this section encompass a range of magnetic field strengths up to 1~T, and initial fast ion energies up to 3.6~MeV. \fgr{taube} presents a scan in energy at three magnetic field strengths. The slowing down time increases with initial fast ion energy in all cases, although slowing down times with and without a background magnetic field are comparable.

\fg[The slowing down time $\tau_{s}$ as a function of initial fast ion energy for a few background magnetic field strengths. The subthermal classical prediction is shown as a solid line.]{taube}{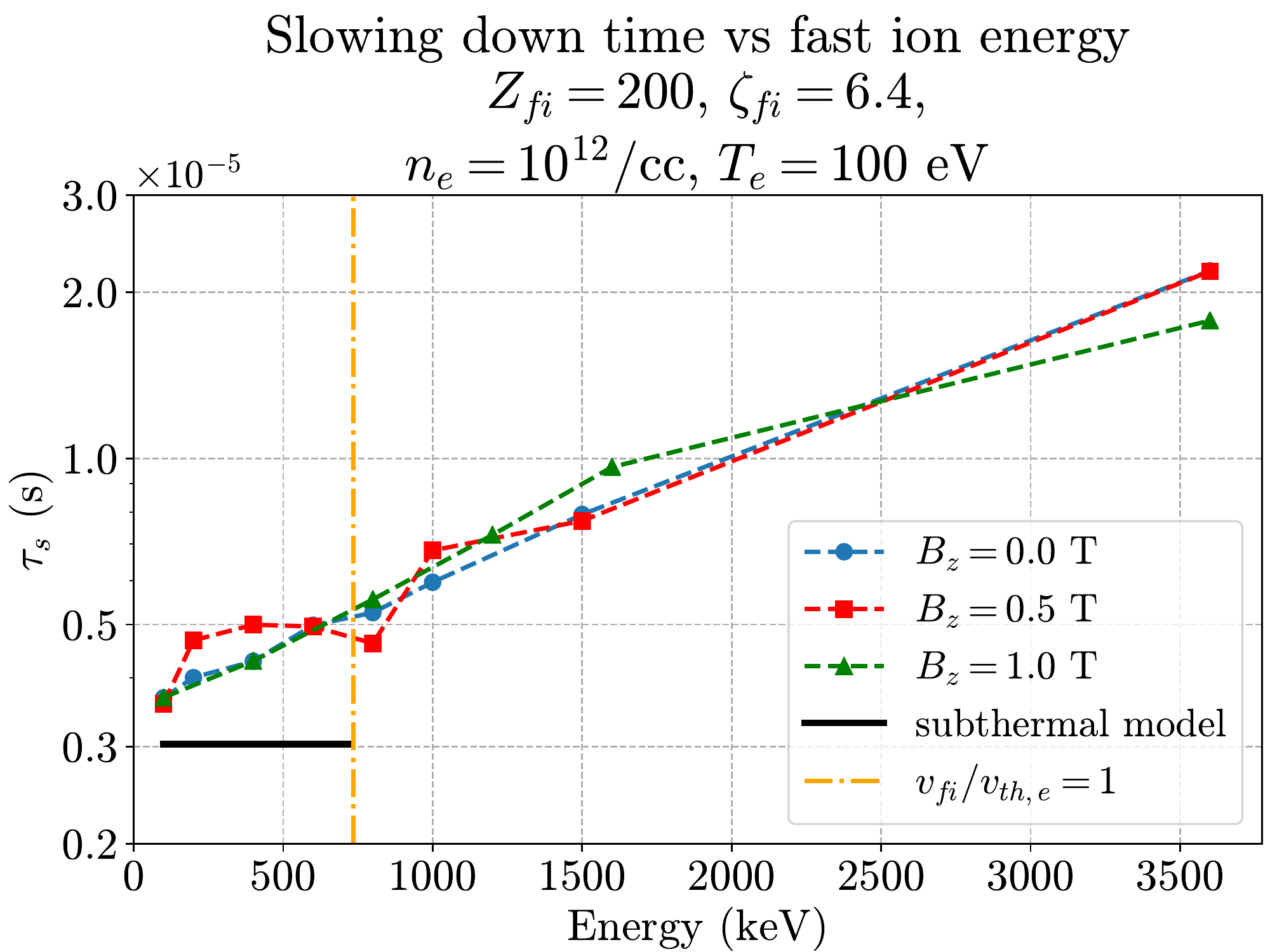}{\figwidth}

\fgr{z2000bscale} presents a higher-resolution scan in magnetic field, with $Z_{fi}=2000$ to decrease the required wall time. Three simulations with different starting particle microstates, for both the background particles and the fast ions, were performed for each data point. The slowing down time is 50\% longer for a field of 1~T as compared to the 0.3~T case, while the 0.5~T case is approximately the same as the 0.3~T case. \fgr{z2000bscaletime} presents aggregate energy time histories for the fast ions at seven of the magnetic field strengths in \fgr{z2000bscale}. Finally, \fgr{z2000parperp} presents a decomposition of the field-parallel and cross-field slowing down rates for the magnetic field scan simulations.

\fg[Detailed scan of background magnetic field, using fast ions at 1~MeV with $Z=2000$. (error bars represent 3 simulations)]{z2000bscale}{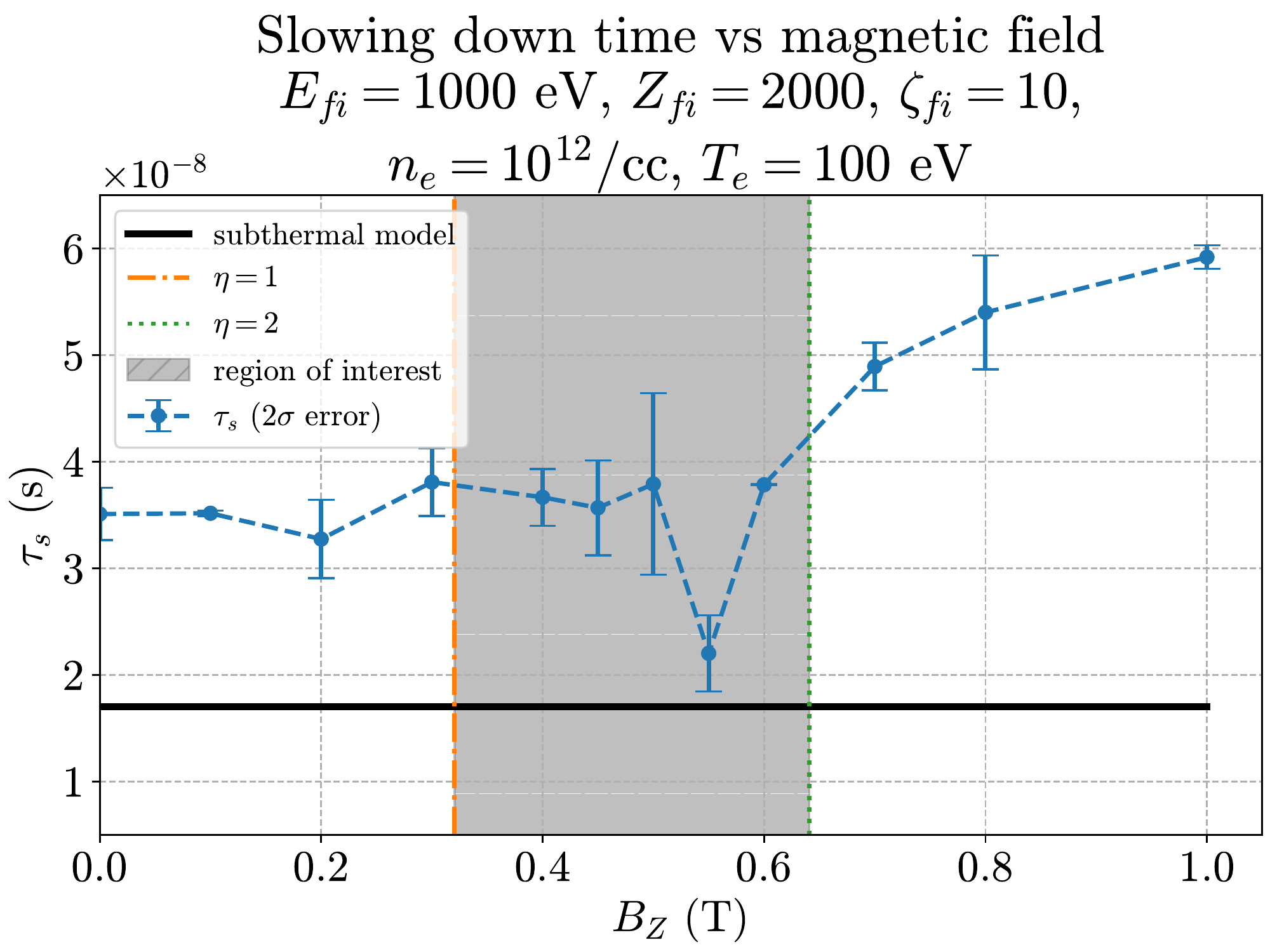}{\figwidth}

\fg[Time histories for one data set from \fgr{z2000bscale}. ]{z2000bscaletime}{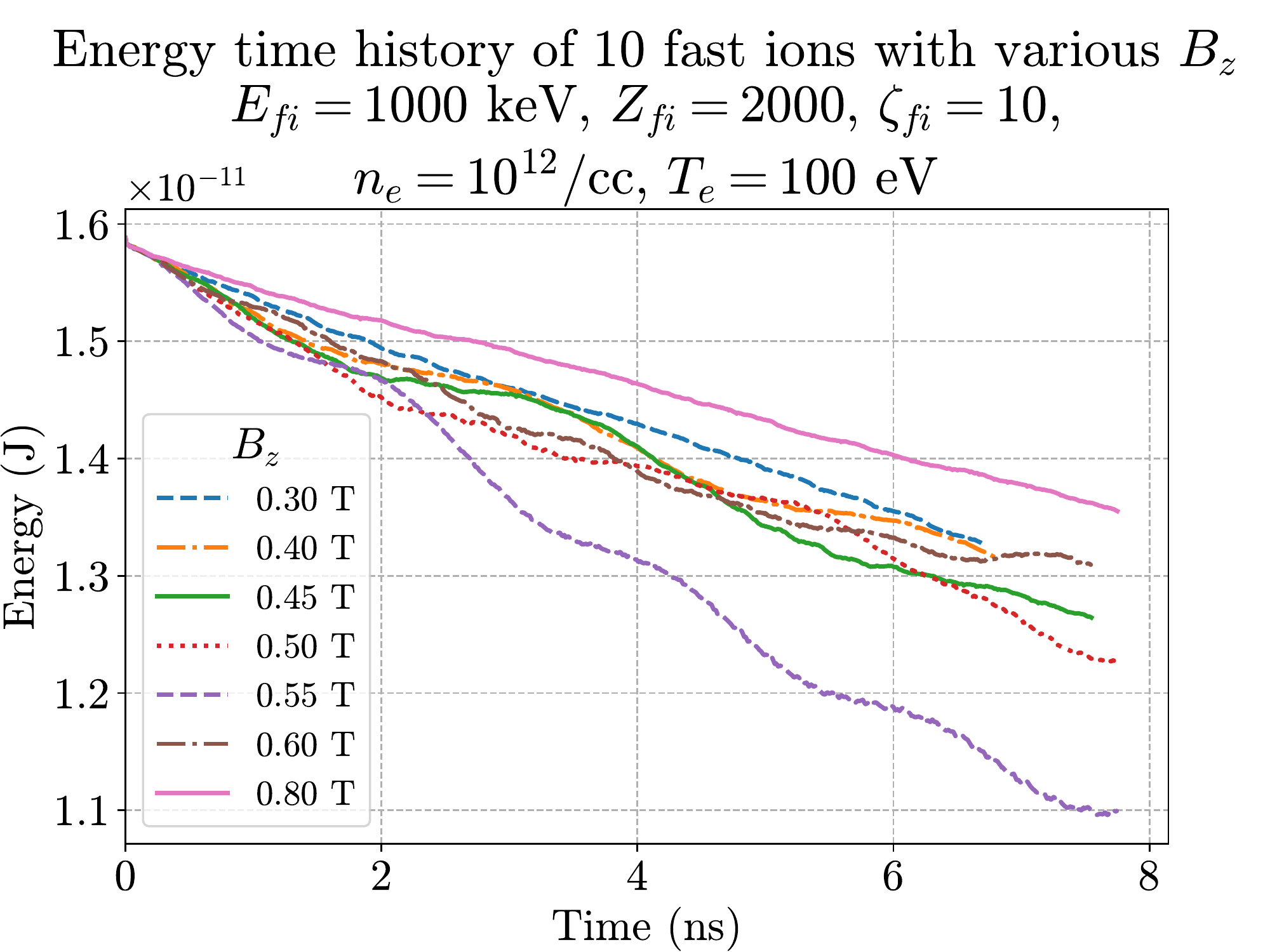}{\figwidth}

\fg[Difference between the cross-field and field-parallel slowing down rates, normalized to the field-parallel slowing down rate (data from \fgr{z2000bscale}); positive values indicate longer cross-field slowing times ($\tau_{\perp}>\tau_{||}$).]{z2000parperp}{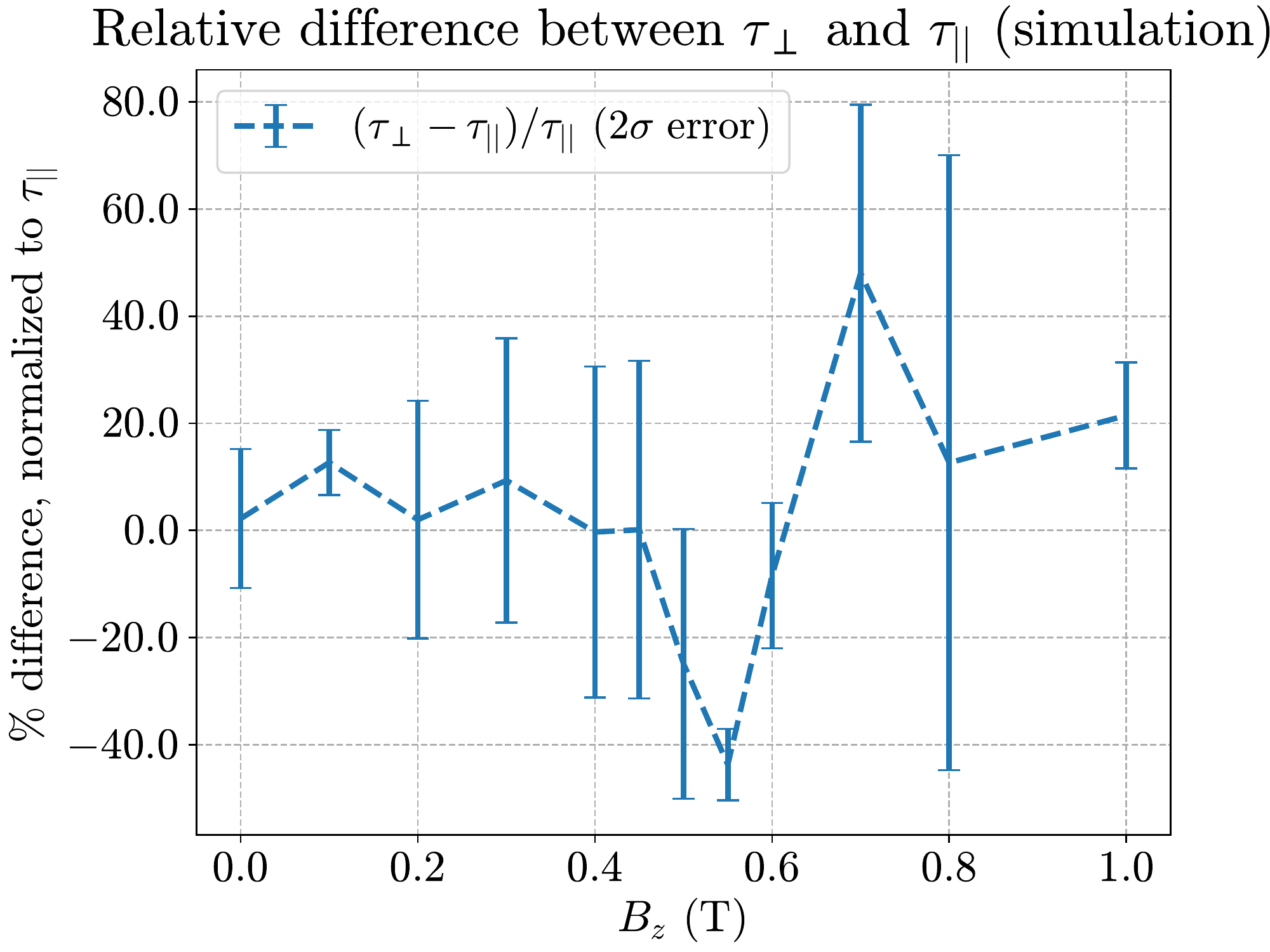}{\figwidth}

\section{Discussion}

The subthermal benchmarking simulations follow the scaling with fast ion charge, $Z_{fi}$, and fast ion clumping factor, $\zeta_{fi}$, as predicted by \eqr{alpha-electron-energy}. For these simulations, the addition of a small background magnetic field ($\rho_{e}/\lambda_{De}=0.64$ for 5~kG and $0.32$ for 10~kG) did not result in a systematic difference with the zero magnetic field case across a range of charges and fast ion energies.

The scaling shown in \fgr{zzetabscale}, an extension of previous work\rfr{kolmes_applications_2014}, allows for an effective speed-up on the order of $10^{6}$, vastly reducing the computational resources required for each simulation. This effect was harnessed to speed up the energy and magnetic field scaling studies involving non-subthermal fast ions.

The excellent agreement between the cell size scan data and the subthermal model when evaluated at the appropriate computational Coulomb logarithm in \fgr{cell-scan-lambda} verifies the formula in \eqr{loglambdasim}. Also, the predictions made by classical slowing down theory are usually only guaranteed to order $\ln\Lambda^{-1}$; however, the slowing-down times in \fgr{cell-scan-lambda} match the classical prediction to within a few percent when the effective computational $\ln\Lambda_{s}\sim 1$.

The magnetic field scan at $Z_{fi}=2000$ (\fgr{z2000bscale}) reveals unexpected structure for intermediate field strengths around 0.5~T. This was not observed to the same extent in simulations with smaller values for the fast ion charge. Possible explanations include nonlinear effects due to high-Z, and/or coupling between the fast ions and waves \rfr{paul_2-d_2013} in the background plasma. The oscillations seen in the fast ion energy time histories from this data (\fgr{z2000bscaletime}) appear to lend some support to the wave hypothesis. Notably, the frequency of the oscillation is approximately equal to the lower hybrid frequency, pointing to possible heating of the background plasma by the fast ions \via waves. Additional simulations concerning this point are on-going.

The cross-field and field-parallel slowing down times in \fgr{z2000parperp} were determined from projections of fast ion trajectories onto the cross-field and field-parallel directions. No consistent anisotropy (with respect to $B_{z}$) is present in this data, and more detailed simulations would be required to reveal the 2-3\% differences predicted in \fgr{hw-diff}.

Although the slowing-down time increases with increasing fast ion energy, the slowing-down time for 1~MeV particles is at most a factor of 2 longer than for 100~keV particles. The addition of a weak magnetic field adds another factor of 1.5 (at $\eta=2$). Thus, the overall increase in slowing-down time for 1~MeV fast ions is a factor of 3 over the subthermal prediction.

\section{Application: FRC Reactor}

Reactor-scale aneutronic (utilizing \deut-\he[3]) FRC designs have been discussed previously \rfr{razin_direct_2014,choi_saffire_1979}. Unlike many MFE reactor designs, the FRC designs referenced here do not rely on any core heating by fusion-produced alphas; all heating power is delivered \via RF. Consequentially, the energy transport focus shifts to the efficient extraction of those alphas and other fusion products, both to remove ash from the core and to recover the produced energy.

Computational studies of fusion product orbits in a reactor-scale FRC using the RMF code\rfr{cohen_ion_2000} have indicated that many fusion products will be born in orbits that traverse the separatrix\rfr{chu_cheong_energetic_2012}. In the cool SOL plasma outside the separatrix, the fusion product experiences orders of magnitude more slowing down than inside. As a result, the fusion product loses energy to the SOL plasma; this ``airbraking'' effect causes the orbit to move out of the core and change character from betatron to cyclotron. The transition from an orbit traversing the separatrix to a cyclotron orbit on only one side of the separatrix is most likely to occur in the SOL, at which point the fusion product is then exhausted along the open field lines of the SOL. This process represents a convenient method of efficiently transporting ash out of the core\rfr{choi_saffire_1979}, facilitating both energy extraction (\via conversion of the heated SOL plasma to electricity) and neutron production mitigation (rapid removal of \trit produced by \deut-\deut reactions).

Using \eqr{tau-energy-subthermal}, $\tau_{s}=0.001$~s for \trit in the SOL (for the conditions of $n_{e}=10^{14}$/cc, $E_{th,e}=100$~eV, $\ln\Lambda=14.7$). Assuming that fusion products spend approximately 10\% of their time in the SOL, and 6 e-folding times (which should drop the triton's energy from 1~MeV to $\sim 20$~keV), the slowing-down time is $t_{s}=0.060$~s. As the average lifetime of \trit is approximately 20~s before fusion, this indicates that the vast majority of \deut-\deut-produced \trit should slow down and be lost along the SOL open field lines rather than fuse. Meanwhile, the combination of the scaling studies for fast ion energy and background magnetic field suggest, overall, an increase in slowing down times of no more than a factor of 2 as compared to the unmagnetized subthermal prediction. Thus, comparing the triton slowing-down to the triton fusion time, we estimate that the neutron production rate should be at least two orders of magnitude lower than in a conventional \deut-\trit tokamak, substantially reducing neutron wall loads and simplifying the engineering requirements for the reactor.

\section{Conclusions}

We have performed PIC simulations of fast ions slowing down on cool weakly magnetized plasma for various macroparticle and charge scaling factors. Our simulations preserve the scalings of those factors predicted by the classical unmagnetized slowing down theory in background plasma with unmagnetized and weakly magnetized electrons. This result indicates that using fast ion macroparticles with hundreds of times more charge than a corresponding real particle can allow relatively short simulations to access slowing down physics on an effectively longer timescale. Specifically, the simulated slowing down rate can be increased by a factor of over $10^{6}$ using fast ion macroparticles with greatly increased charge, allowing millisecond-scale slowing down physics to be accessed by a nanosecond-scale simulation. Simulations of MeV-scale fast ions (with $\xi_{fi}\equiv v_{fi}/v_{th,e}>1$) indicate that the unmagnetized subthermal slowing down prediction should be increased by no more than a factor of 3 to account for the combined effects of superthermal fast ions and a weak magnetic field ($1<\eta\equiv\Omega_{ce}/\omega_{pe}<2$). This result indicates that fusion products in a small aneutronic FRC reactor can be quickly exhausted \via slowing down in the SOL, for efficient power and ash extraction as well as neutron production mitigation.



%
%

%

\begin{acknowledgments}
We thank Dr. G.~Hammett and C.~Swanson for useful suggestions, and the Program in Plasma Science and Technology for support under contract number DE-AC02-09CH11466. Early contributions to this study were made by  A.~Creely, E.~Paul and E.~Kolmes. Most simulations were performed on the Cori, Edison and Hopper clusters at NERSC. Most plots were produced with Python3/Matplotlib\rfr{hunter_matplotlib:_2007}.
\end{acknowledgments}

\bibliography{esevans-slowing-down-aip-pop-final}

\end{document}